\documentstyle[prd,aps,psfig]{revtex}
\newcommand{\lapproxeq}
{\lower .7ex\hbox{$\;\stackrel{\textstyle <}{\sim}\;$}}
\begin{document}
\draft
\input epsf
\twocolumn[\hsize\textwidth\columnwidth\hsize\csname
@twocolumnfalse\endcsname
%
%
\title{{\hfill  \rm \normalsize DSF-14/2001, astro-ph/0105385}\\ $~$\\
Constraining neutrino physics with BBN and CMBR}

\author{S. H. Hansen${}^{1}$, G. Mangano${}^{2}$,
A.\ Melchiorri${}^{1}$, G.\ Miele${}^{2}$, and O.\ Pisanti${}^{2}$}
\address{${}^1$ Department of Physics, Nuclear \& Astrophysics Laboratory,
University of Oxford, Keble Road, Oxford OX1 3RH, U.K.}
\address{${}^2$ Dipartimento di Scienze Fisiche, Universit\`{a} ``Federico
II'', Napoli, and INFN Sezione di Napoli, Complesso Universitario di Monte
Sant'Angelo, Via Cintia, 80126 Napoli, Italy.}
\maketitle

\begin{abstract}
{We perform a likelihood analysis of the recent results on the anisotropy
of Cosmic Microwave Background Radiation from the BOOMERanG and DASI
experiments to show that they single out an effective number of neutrinos
in good agreement with standard Big Bang Nucleosynthesis. We also consider
degenerate Big Bang Nucleosynthesis to provide new bounds on effective
relativistic degrees of freedom $N_\nu$ and, in particular, on neutrino
chemical potential $\xi_\alpha$. When including Supernova Ia data we find,
at $2\sigma$, $N_\nu \leq 7$ and $-0.01 \leq \xi_e \leq 0.22$,
$|\xi_{\mu,\tau}|\leq 2.6$.}
\end{abstract}
\pacs{PACS number(s): 98.80.Cq, 04.50.+h, 95.35.+d, 98.70.Vc}

\vskip2pc]

\def\gsim{\;\raise0.3ex\hbox{$>$\kern-0.75em\raise-1.1ex\hbox{$\sim$}}\;}
\def\lsim{\;\raise0.3ex\hbox{$<$\kern-0.75em\raise-1.1ex\hbox{$\sim$}}\;}

\section{Introduction}
\label{secintobs}

New results on Cosmic Microwave Background Radiation (CMBR) anisotropy from
BOOMERanG \cite{Boom1,Boom2}, MAXIMA \cite{Max1,Max2}, and DASI \cite{Dasi}
experiments represent an extraordinary confirmation of our present
understanding of some of the key features of the evolution of our universe.
The clean evidence for the first acoustic peak of temperature anisotropies
for CMBR around $l \sim 200$ \cite{Boom1,Boom2,Max1,Max2,Dasi,stompor}
strongly supports the scenario of a post-inflationary flat universe. On
the other hand new results on the second and third peak confirm the
adiabatic inflationary model prediction of acoustic oscillations in the
primeval plasma driven by gravity, and shed new light on how energy
density is distributed among several components. This is a crucial piece
of information which affects many independent cosmological observables, so
it is reasonable to expect that it will be possible in the next years to
have a rather clear picture of which cosmological model is actually
realized in our universe.

In this respect BOOMERanG and MAXIMA first data release \cite{Boom1,Max1}
already stimulated a wide number of studies
\cite{Max1,th1,th3,th4,th4b,th4bb,th5,th6,th7,th8}, aimed to constrain the
values of the energy density parameters normalized to the critical density,
$\Omega_b$, $\Omega_m$ and $\Omega_\Lambda$, due to baryons, dark matter
and an effective cosmological constant, respectively. In particular many
authors have addressed the issue of a  tension between the determination of
$\Omega_b h^2$ from CMBR data and Standard Big Bang Nucleosynthesis (SBBN)
\cite{th4,th5,th7,Emmp2,Hansen,Kaplin,dibari,alpha}. In fact, the finding
of a suppressed second peak in the CMBR anisotropy resulted in a rather
large value for this parameter, $\Omega_b h^2 = 0.032^{+0.005}_{-0.004} $
at $68\%$ CL~\cite{th1},  while the experimental data on primordial $^4He$
and $D$ abundances, prefer smaller values, $\Omega_b h^2 =
0.019^{+0.004}_{-0.002}$, Ref. \cite{alpha}, and $\Omega_b h^2 = 0.020
{\pm} 0.002$, Ref. \cite{Burles} (see also \cite{Lisi}), at $95 \%$ CL.
These estimates are obtained assuming three standard neutrino degrees of
freedom.

New experimental data from BOOMERanG has refined the data at larger
multipoles, and now single out a smaller value for the baryonic fraction,
$\Omega_bh^2= 0.021^{+0.004}_{-0.003}$ \cite{Boom2}. This is mainly due to
an increase in the analyzed dataset (roughly by a factor 8) and a better
understanding of the experimental beam, calibration and pointing. The new
analysis leads to a slightly increased amplitude for the second peak (but
still compatible with the previous spectrum) and hints for the presence of
a third peak around $l\sim 850$, which is not as high as expected in a
scenario with a large baryonic fraction. Simultaneously the DASI
experiment, which also found evidence for multiple peaks in the CMBR
spectrum, gave an impressive and independent confirmation of a low baryon
fraction, $\Omega_bh^2 = 0.022^{+0.004}_{-0.003}$ \cite{Dasi}, when
sampling a different region of the sky and different frequencies. It is
worth stressing that these high multipole data may still be affected by
large systematic errors (see for example the consistency test in Table $3$
in Ref. \cite{Boom2}), thus all conclusions relying on them should still be
taken with caution. This is especially true in view of the revised spectrum
at $\ell \ge 300$ from the Maxima-I experiment, which gives the wide range
$\Omega_bh^2 = 0.0325 {\pm} 0.0125$ \cite{stompor}.

Nevertheless it is important, on the basis of the new data now available,
to undertake a detailed study of the compatibility of these data with SBBN.
For this purpose we have performed, as in~\cite{th7}, a likelihood analysis
of BOOMERanG/DASI CMBR data and SBBN in the parameter space $(\Omega_b
h^2$, $N_\nu)$, with $N_\nu$ the effective neutrino degrees of freedom, and
indeed we find a very good agreement. In particular the SBBN $95 \%$ CL
region, corresponding to $N_\nu = 2.8 {\pm} 0.3$ and $\Omega_b h^2= 0.020
{\pm} 0.004$, has a large overlap with the analogous CMBR contour. This
fact, if it will be confirmed by future experiments on CMBR anisotropy,
can be seen as one of the greatest success, up to now, of the standard hot
big bang model.

As a byproduct of our analysis we also comment on the possible primordial
$^7Li$ depletion, which has already been discussed in the literature
\cite{Burles,Ryan,Salaris}. We find that a depletion factor $f_7 \sim 1/2
\div 1/3$ may reconcile observations from Spite plateau with the value of
$\Omega_b h^2$.

SBBN is well known to provide strong bounds on $N_\nu$. On the other hand,
Degenerate BBN (DBBN), first analyzed in Ref. \cite{d1,d2,d3,Kang}, gives
very weak constraint on the effective number of massless neutrinos, since
an increase in $N_\nu$ can be compensated by a change in both the chemical
potential of the electron neutrino, $\mu_{\nu_e}= \xi_e T_\nu$, and
$\Omega_bh^2$. However, combining this scenario with the bound on baryonic
and radiation densities allowed by CMBR data, it is possible to obtain
rather strong constraints on $N_\nu$ even for DBBN. From our analysis we
get the bound $N_\nu \leq 7$, at $95 \%$ CL, when including Supernovae Ia
(SNIa) data, which translates into a new and more stringent bound on
background neutrino chemical potentials.

Some caution is naturally necessary when comparing the effective number of
neutrino degrees of freedom from BBN and CMBR, since they may be related to
different physics. In fact the energy density in relativistic species may
change from the time of BBN ($T \sim MeV$) to last scattering ($T \sim
eV$). Specifically, if a neutrino has a mass in the range $eV < m < MeV$,
and decays into sterile particles, like other neutrinos, majorons etc.,
with lifetime $t(\mbox{BBN}) < \tau < t(\mbox{CMBR})$, then the effective
number of neutrinos at CMBR would be sensibly different than at BBN
\cite{White:1995as}. However, this possibility does not look too natural
any longer, in view of the recent experimental results on neutrino
oscillation \cite{Fogli,Valle}, showing that all active neutrinos are
likely to have masses smaller than $eV$. One could instead consider sterile
neutrinos mixed with active ones, which could be produced in the early
universe by scatterings and subsequently decay. However, for mixing angle
large enough to thermalize sterile neutrinos \cite{langacker}, one needs a
sterile to active neutrino number density ratio $n_s / n_\nu \approx 4
{\cdot} 10^4 \sin^2 2\theta \ (m/\mbox{keV}) (10.75/g^*)^{3/2}$ of order
unity \cite{DH} ($\theta$ is the mixing angle, and $g^*$ is the number of
relativistic degrees of freedom). Hence using the decay time, $\tau
\approx 10^{20} (\mbox{keV}/m)^5 /\sin^2 2\theta$ sec, one finds $\tau
\approx 10^{17} (keV/m)^4 \, yr$, which is much longer than the age of the
Universe, so they would certainly not have decayed at $t(\mbox{CMBR})$.
Seemingly a sterile neutrino with mass of few MeV would have the right
decay time, but this is excluded by standard BBN
considerations~\cite{massive,massive2}. Let us emphasize that even though
the simplest models allow to directly combine BBN and CMBR results,
nevertheless one may consider more exotic scenarios
\cite{Bartlett:1991qq,Kaplin}, where $\Omega_b h^2$ changes between BBN and
CMBR epochs, or quintessence, which would result in a change of $N_\nu$
between BBN and CMBR \cite{Bean:2001wt}.

The paper is organized as follows: Section II is devoted to a brief review
of the data used in our analysis, which is contained in Section III.
Finally in Section IV we give our conclusions.

\section{BBN and CMBR data}
\label{bbn}

A faithful estimate of primordial Deuterium is provided by Ly-$\alpha$
features in several Quasar Absorption Systems (QAS) at high red-shift ($z
\geq 2$). The most recent analysis of a four QAS sample gives $D/H = (3.0
{\pm} 0.4){\times}10^{-5}$ \cite{Tytler}. A new measurement has been also
presented from observations of the Q-2206-199 QAS, at red-shift $z \simeq
2$, which gives $D/H = (1.65 {\pm} 0.35){\times}10^{-5}$ \cite{Pettini}.
When combined with the data of Ref. \cite{Tytler} this result gives a
sensibly lower estimate for D abundance, $D/H =(2.2 {\pm}
0.2){\times}10^{-5}$. Nevertheless, as reference value we will use the
result quoted in \cite{Tytler}, but we will comment in our final
discussion on the possible impact of what was found in \cite{Pettini} in
the determination of $\Omega_b h^2$ from BBN and CMBR new data.

For the $^4He$ mass fraction, $Y_P$, the key results come from the study of
HII regions in Blue Compact Galaxies. The most complete and homogeneous
sample has been analyzed in Ref. \cite{IzotovThuan}, giving the value $Y_P
= 0.244 {\pm} 0.002$. A recent study, however, has pointed out the presence
of possible systematic errors in inferring the total $^4He$ abundance due
to both imperfect ionization and non uniform temperature distribution
\cite{Sauer}, leading to a typical overestimation of $(2 \div 4 )\%$ of
$Y_P$. This issue of course deserves a deeper study to understand if
uncertainties in $^4He$ measurements are actually dominated by systematic
effects. Notice that in the extreme case, a value as low as $Y_P= 0.234$
may represent a new problem for the very consistency of BBN scenario, in
view of the low $D$ result of \cite{Tytler}. In what follows we will use
with caution the result of Ref. \cite{IzotovThuan} quoted above.

The estimate of $^7Li$ primordial abundance using Spite plateau can be
spoiled by four possible systematic effects \cite{Ryan}: a) Galactic
Chemical Evolution (GCE), which is poorly known; b) corrections for
possible depletion of initial star surface abundance; c) the very method of
how $^7Li$ is obtained from Spite plateau; d) presence of anomalous stars
in the samples. In particular the effect due to GCE was long assumed to be
negligible for metal poor stars in view of its apparent uniformity, but
this has recently been questioned due to observation of some amount of
$Be$. Furthermore, data shows a statistically significant increase with
$Fe/H$, as shown in \cite{Ryan}, leading to a primordial Lithium abundance
$^7Li/H = (1.23^{+0.68}_{-0.32}) {\times} 10^{-10}$. Evidence for this
effect was instead missing in a previous analysis \cite{Li7a}, where it
was found $^7Li/H = (1.73 {\pm} 0.21) {\times} 10^{-10}$. The effects b)
and c) have also recently been studied in \cite{Salaris}, where it is
pointed out that Spite plateau can be well reproduced by models with a
strong diffusion effect, and would be a factor two lower than the
primordial abundance.

For these reasons, at present it is not appropriate to include $^7Li$ in a
likelihood analysis of BBN. As in \cite{Burles}, we will rather estimate
from BBN prediction the depletion factor $f_7$=$^7Li_{obs}/^7Li_{prim}$,
using as a reference result the one quoted in \cite{Ryan}.

The anisotropy power spectrum from BOOMERanG experiment was estimated in
$19$ bins between $\ell$ $ =75$ and $\ell=1025$. Since the correlation
matrix still is not public available, we will assume the data points to be
independent. The data provide evidence for the presence of $3$ peaks at
$\ell \sim 210_{-9}^{+5}$, $550_{-12}^{+8}$, $840_{-13}^{+6}$, with an
amplitude of $\sim 72 \mu K$, $49 \mu K$ and $45 \mu K$ respectively
\cite{paolo2}.  In our analysis we include a $10 \%$ correlation between
the signal $C_B$ in the bin $B$, a calibration uncertainty of $25 \%$ in
$\Delta T^2$ and a gaussian uncertainty of $1.4'$ in the beam. Furthermore,
since the signal at very high multipoles ($\ell \ge 850$) could be severely
affected by the presence of systematic effects, we apply a jackknife test
repeating the analysis without these datapoints, finding no significant
changes in our results. For the DASI data we include the window functions
available on the corresponding web site \cite{dasiweb}. We also include a
$8 \%$ calibration error. There is an $\sim 20 \%$ overlap of the two
regions of the sky covered by the two experiments but we do not take this
effect into account in our analysis. In fact we believe that this
correlation should not affect our conclusions, since our result appears
stable when removing the DASI data points.

\section{Likelihood analysis}

The likelihood analysis of the BBN data has been performed using the method
already described in details in \cite{Emmp2}. To constrain the values of
the parameter set $(N_\nu,\Omega_b h^2)$, for SBBN, and $(\xi_e, N_\nu,
\Omega_b h^2)$, for the degenerate scenario, from the data on $^4He$ and
$D$, we define the likelihood function ${\cal L}_{BBN}$ = $L_D ~L_{^4He}$,
where each likelihood function, assuming gaussian distribution for the
errors, is given by the overlap of a theoretical and experimental
distribution,
\begin{eqnarray}
&L_i =& \frac{1}{2\, \pi\, \sigma_i^{th} \, \sigma_i^{ex}}
{\cdot}\label{likei} \\
& {\cdot} & \int dY~ \exp \left\{ - \frac{(Y-Y_i^{th})^2}{2\,
\sigma_i^{th\, 2} } \right\} \exp \left\{ - \frac{(Y-Y_i^{ex})^2}{2\,
\sigma_i^{ex\, 2}} \right\} \, .
 \nonumber
\end{eqnarray}
The $Y_i^{ex}$ and $\sigma_i^{ex}$ are the experimental results and
$1-\sigma$ errors for the i-th nuclide, $Y_i^{th}$ the theoretical
predictions obtained by an updated BBN code developed over the last few
years \cite{Emmp1,Emmp2}. Finally, the theoretical $\sigma_i^{th}$ can be
found by linear propagation of the uncertainties of the various nuclear
rates entering in the nucleosynthesis reaction network \cite{Lisi}.

For the BOOMERanG and DASI experiments we approximated the likelihood
function of the CMBR signal inside the bins, $C_B$, as a gaussian variable.
The likelihood for a given cosmological model is then defined by $-2{\rm
ln} {\cal L}_{CMBR} =(C_B^{th}-C_B^{ex})^2/\sigma_B^2$, where $C_B^{th}$
is the theoretical signal. Our database of models is sampled as in
\cite{th7}.

As can be seen from Figure 1, the dotted (red) line, which represents the
$95 \%$ CL contour of SBBN, is in very good agreement with new CMBR data,
and the $\Omega_b h^2$ tension between primordial nucleosynthesis and CMBR
anisotropy seems to be completely solved. The constraint on $\Omega_b h^2$
($N_\nu$) can be obtained by marginalizing the total likelihood function
${\cal L}={\cal L}_{SBBN}{\cdot}{\cal L}_{CMBR}$ with respect to $N_\nu$
($\Omega_b h^2$). By this procedure we get the two estimates $\Omega_b h^2
= 0.019{\pm} 0.003$ and $N_\nu = 2.8 {\pm} 0.4$, both at $95 \%$.

The result on $N_\nu$ beautifully suggests the simplest scenario of three
light active neutrinos. It is therefore perfectly meaningful to fix from
the very beginning
$N_\nu=3.034$~\cite{Dicus,Dolgov:1997sf,Dolgov:1999sf,EMPPP}, which leads
to the same interval for $\Omega_b h^2$. In particular, for $\Omega_b h^2 =
0.019$ the nuclei abundances evaluate to $D/H = 3.26{\times}10^{-5}$, $Y_P=
0.2471$ and $^7Li/H= 3.31{\times}10^{-10}$.

\begin{center}
\begin{figure}
\epsfxsize=8.cm
\epsfysize=8.cm
\epsffile{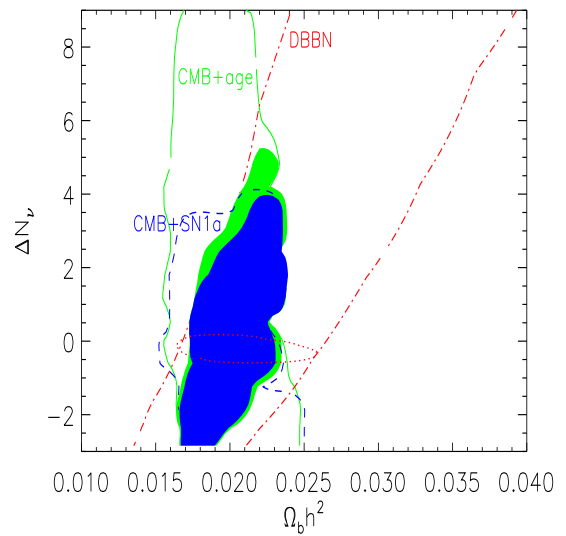}
\caption{The $95\%$ CL contours for degenerate BBN (dot-dashed (red) line),
new CMBR results only with age prior, $t>11$gyr (full (green) line), and
only with SNIa prior (dashed (blue) line) are shown. The combined analyses
correspond to filled areas: DBBN + CMBR + age (light (green) region), DBBN
+ CMBR + SNIa (dark (blue) region). The dotted (red) line is the $95 \%$ CL
contour of SBBN.}
\label{figDbbn}
\end{figure}
\end{center}

Although SBBN seems to be in very good agreement with the new CMBR data,
providing a very strong constraint on the neutrino degrees of freedom, it
relies on the theoretical assumption that background neutrinos have
negligible chemical potential, as their charged lepton partners. Even
though this hypothesis is perfectly justified by Occam razor, models have
been proposed in the literature
\cite{dibari,AF,DK,DolgovRep,Casas,MMR,McDonald,Foot} where large neutrino
chemical potentials can be generated. It is therefore an interesting issue
for cosmology, as well as for our understanding of fundamental
interactions, to try to constraint the neutrino--antineutrino asymmetry
from cosmological observables. It is well known that degenerate BBN gives
severe constraints on the electron neutrino chemical potential, $-0.06\leq
\xi_e\leq 1.1$, and weaker bounds on the ones of both $\mu$ and $\tau$
neutrino, $|\xi_{\mu,\tau}| \leq 5.6 \div 6.9$ \cite{Kang}. This occurs
since electron neutrinos are directly involved in neutron to proton
conversion processes which eventually fix the total amount of $^4He$
produced in nucleosynthesis, while $\xi_{\mu,\tau}$ only enters via their
contribution to the expansion rate of the universe. Combining this
scenario with the bound on baryonic and radiation densities allowed by
CMBR data, it is possible to obtain rather stronger constraints on all
these parameters. Such an analysis was previously performed in
\cite{th7,steig} using BOOMERanG and MAXIMA data of Refs.
\cite{Boom1,Max1}. We recall that neutrino chemical potentials contribute
to the total neutrino effective degrees of freedom $N_\nu$ as
\begin{equation}
N_{{\nu}} = 3 + \Sigma_{\alpha} \left[ \frac{30}{7} \left(
\frac{\xi_\alpha}{\pi} \right)^2 + \frac{15}{7} \left(
\frac{\xi_\alpha}{\pi} \right)^4 \right] \, + \delta_\nu\, ,
\end{equation}
with $\delta_\nu$ the contribution of relativistic degrees of freedom other
than neutrinos and photons. Notice that, in order to get the most stringent
bound on $\xi_\alpha$ we have to assume that all relativistic degrees of
freedom, other than photons, are given by three active (possibly)
degenerate massless neutrinos, i.e. $\delta_\nu=0$. Similarly the upper
limit on $\delta_\nu$ can be obtained from the results of our analysis in
the case $\xi_{\mu,\tau}=0$. We stress that in any case a value for $N_\nu$
sensibly different than three $does$ require a non vanishing chemical
potential for electron neutrinos, or more generally a non thermal spectrum.

Figure 1 summarizes our main results for the DBBN scenario. Defining
$\Delta N_\nu=N_\nu-3$, we plot in the plane ($\Delta N_\nu, \Omega_b h^2)$
the $95 \%$ CL contour allowed by DBBN (dot-dashed (red) line), together
with the analogous $95 \%$ CL region coming from the CMBR data analysis,
with only weak age prior, $\tau > 11 $gyr (full (green) line). Finally,
the light (green) filled region is the $95 \%$ CL region of the joint
product distribution ${\cal L} \equiv {\cal L}_{DBBN}$${\cdot}{\cal
L}_{CMBR}$. The main new feature, with respect to the results of Ref.
\cite{th7} is that the resolution of the third peak shifts the CMBR
likelihood contour towards smaller values for $\Omega_b h^2$, so, when
combined with DBBN results, it singles out smaller values for $N_\nu$. In
fact from our analysis we get the bound $N_\nu \leq 8$, at $95 \%$ CL,
which translates into the new bounds $-0.01\leq \xi_e \leq 0.25$, and, for
$\delta_\nu=0$, $|\xi_{\mu,\tau}| \leq 2.9$, sensibly more stringent than
what can be found from DBBN alone.

A similar analysis can be also performed combining CMBR and DBBN data with
the Supernova Ia data \cite{sn1a}, which strongly reduce the degeneracy
between $\Omega_m$ and $\Omega_\Lambda$. At $95 \%$ CL we find (dark (blue)
filled region in Figure 1) $N_\nu \leq 7$, corresponding to  $-0.01\leq
\xi_e \leq 0.22$ and $|\xi_{\mu,\tau}|\leq 2.6$, for $\delta_\nu=0$. In
the other extreme scenario, where basically all extra contributions to
Hubble parameter are given by extra relativistic species, we get
$\delta_\nu \leq 4$.

Another possibility to break the degeneracy between $\Omega_m$ and
$\Omega_\Lambda$, is to put priors on the age of the universe $\tau$, as
pointed out in Ref. \cite{steig}.  In Figure 2 we show the normalized
likelihood functions for age priors of $\tau>10,11,..,14$ gyr, using only
CMBR data. It is clear that one needs the slightly unrealistic prior of
$\tau>13$ gyr to get bounds stronger than $\Delta N \leq 4$, as obtained by
the inclusion of SNIa data.

\begin{center}
\begin{figure}
\epsfxsize=8.cm
\epsfysize=8.cm
\epsffile{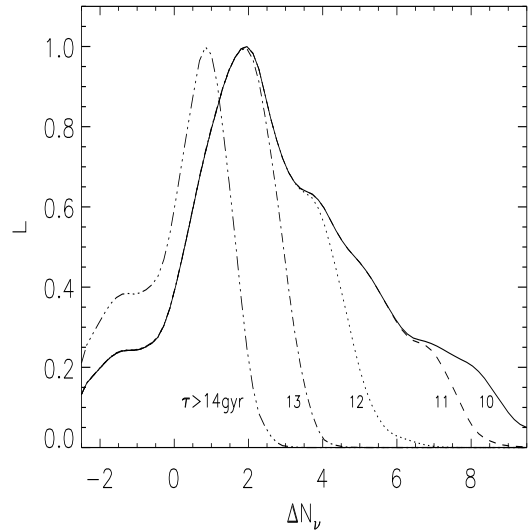}
\caption{CMBR likelihoods for age priors of $\tau>10,11,..,14$ gyr.}
\label{ageprior}
\end{figure}
\end{center}

Recently Ref.~\cite{shd} stressed the point that the inclusion of large
scale structure data can provide a lower bound on $\Delta N_\nu$. In
Ref.~\cite{shd} SNIa data are not considered, neither a DBBN scenario;
however, where comparison is possible, the results on upper bounds are in
fair agreement with our results\footnote{A study of the effects of large
neutrino asymmetries on CMBR + large scale structures has also been
performed in Ref. \cite{LL}, where the compatibility with DBBN for $m_\nu
\lapproxeq 1 eV$ has been analyzed. Their results refer to a critical
universe with no cosmological constant.}. It is worth noticing once again
that our rather stringent bound on $N_\nu$ is the outcome of the combined
analysis of CMBR $and$ DBBN. Each of the two corresponding likelihood
contours in fact, taken separately, give a much weaker bound (see Figure
1).

As we mentioned in Section 2, it has recently been stressed that depletion
effects on $^7Li$ may be efficient in reducing the primordial abundance
down to the value observed in Spite plateau. In Figure 3 we plot the $^7Li$
depletion factor $f_7$, defined as the ratio of the experimental value of
Ref. \cite{Ryan} and the theoretical estimate from our BBN code. Values for
$f_7$ of the order of $1/2 \div 1/3$ cannot be ascribed to a statistical
fluctuation in the star sample considered in \cite{Ryan}, but should rather
be understood by a careful analysis of all systematic effects which we
briefly reviewed in Section II.

There are some points we would like to address as final remarks. First of
all we stress once again that further data on the third peak in the CMBR
anisotropy spectrum are needed to check for possible systematics. This is a
crucial point for a clean determination of the baryonic fraction, since
discrimination between SBBN and DBBN, or SBBN and other theoretical
framework for light nuclei production, relies on both second {\it and }
third peak heights. In this respect we note the good agreement between the
BOOMERanG and DASI results.

As a second observation, we recall that we already pointed out that a new
measurement of primordial $D$ has been reported recently, leading to a
weighted average $D/H =(2.2 {\pm} 0.2){\times}10^{-5}$ \cite{Pettini}. If
we adopt this different estimate, the overlap of SBBN and CMBR contours
decreases, but still there is a good agreement at $95 \%$ CL. Using the
SBBN likelihood analysis only we find in this case, at $95 \%$ CL,
$\Omega_b h^2 = 0.024^{+0.004}_{-0.003}$ and $N_\nu=2.7{\pm}0.3$, while
combining this result with CMBR data and using the joint likelihood
distribution ${\cal L}$ we get $\Omega_b h^2=0.023^{+0.005}_{-0.003}$ and
$N_\nu=2.7{\pm}0.3$.

\begin{center}
\begin{figure}
\epsfxsize=8.cm
\epsfysize=7.cm
\epsffile{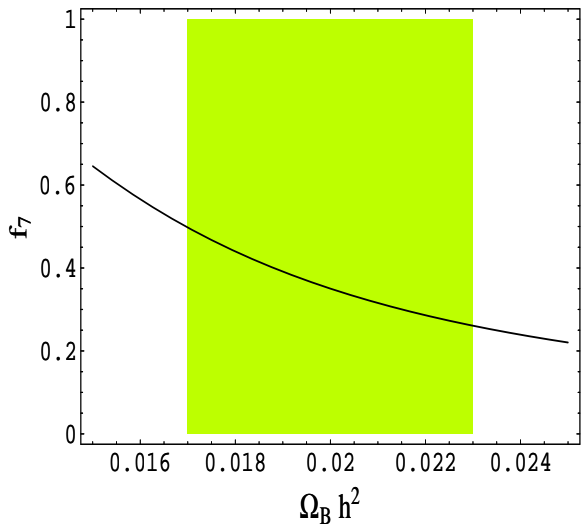}
\caption{The $^7 Li$ depletion factor, defined as the ratio between the
experimental and theoretical values.}
\label{figli7}
\end{figure}
\end{center}

\section{Conclusions}
\label{secdiscon}

It is a great success of cosmology and astrophysical observations that
severe constraints can be put on the number of neutrino degrees of freedom
and, more generally, of light particle species which were relativistic at
the epoch of recombination. Of course this is a fundamental piece of
information for the whole microscopic theory of fundamental interactions.
The increasing precision in measurements of primordial abundances of light
nuclei, and the impressive progress in measuring the CMBR anisotropy, are
conspiring to give us a very precise determination of $N_\nu$. Despite of
the conservative expectation of three, light, active neutrinos, largely
non-degenerate, it should be stressed that many other scenarios have been
considered in the literature, based on theoretical ideas which, going
beyond the Standard Model, try to grasp possible extension of our
knowledge of fundamental interactions at higher energy. It is really
exciting that, along with customary accelerator physics, we have at hand a
severe way to scrutinize these models by cosmological measurements.

In this paper we have studied in details the implications of the new
BOOMERanG and DASI data on CMBR anisotropy for the estimation of the
baryonic energy density fraction, compared with the predictions of standard
BBN, in the parameter space $\Omega_b h^2-N_\nu$. Observation of the third
peak at multipole $l \sim 850$ turned into a sensible improvement of the
compatibility of the two independent ways of constraining $\Omega_b$, and
single out the values $\Omega_b h^2=0.019{\pm} 0.003$, and $N_\nu = 2.8
{\pm} 0.4$, both at $2\sigma$.

We have also considered the scenario of a degenerate neutrino background,
which strongly affects primordial nuclei production. The new CMBR BOOMERanG
and DASI data lead to a new and stronger constraint on the effective
relativistic degrees of freedom, $N_\nu \leq 8$ (only weak age prior), or
$N_\nu \leq 7$ (with only SNIa prior), both at $95 \%$ CL, which bounds
more severely the neutrino chemical potentials, $|\xi_{\mu,\tau}| \leq
2.9$, and $-0.01 \leq \xi_e \leq 0.25$, and $|\xi_{\mu,\tau}| \leq 2.6$,
and $-0.01 \leq \xi_e \leq 0.22$, respectively.

\section*{Acknowledgement}
SHH is supported by a Marie Curie Fellowship of the European Community
under the contract HPMFCT-2000-00607. AM is supported by PPARC.


\end{document}